\documentclass{llncs}
\usepackage{mathptmx}
\usepackage{moreverb}
\usepackage{color}
\usepackage{graphicx}
\usepackage{listings}
\usepackage{array}
\usepackage{times}
\usepackage{rotating}
 \usepackage[font={small}]{caption, subfig}
\newcommand{\bq}{\begin{equation}}
\newcommand{\eq}{\end{equation}}

\newcommand{\flops}{\mbox{flops}}
\newcommand{\update}{\mbox{update}}
\newcommand{\updates}{\mbox{updates}}
\newcommand{\UP}{\mbox{UP}}
\newcommand{\GUS}{\mbox{G\UP/\second}}

\newcommand{\second}{\mbox{s}}

\newcommand{\FS}{\mbox{F/s}}
\newcommand{\GBS}{\mbox{GB/s}}

\newcommand{\GHZ}{\mbox{GHz}}
\newcommand{\GCS}{\mbox{G\cycles/\second}}

\newcommand{\bytes}{\mbox{bytes}}
\newcommand{\byte}{\mbox{byte}}
\newcommand{\BYTE}{\mbox{B}}

\newcommand{\MiB}{\mbox{MiB}}

\newcommand{\cycles}{\mbox{cy}}

\newcommand{\eos}{~.}

\newcommand{\Rlm}{Roof\/line model}
\newcommand{\ecmm}{ECM model}

\newcommand{\construction}[1]{}

\newcommand{\olsep}{\|}
\newcommand{\nolsep}{|}
\newcommand{\ecmspace}{\,}
\newcommand{\ecm}[6]{\mbox{$\left\{{#1}\ecmspace\olsep\ecmspace {#2}\ecmspace\nolsep\ecmspace {#3}\ecmspace\nolsep\ecmspace {#4}\ecmspace\nolsep\ecmspace {#5}\right\}\ecmspace{#6}$}}
\newcommand{\epsep}{\rceil}
\newcommand{\ecmp}[5]{\mbox{$\left\{{#1}\ecmspace\epsep\ecmspace {#2}\ecmspace\epsep\ecmspace {#3}\ecmspace\epsep\ecmspace {#4}\right\}\ecmspace{#5}$}}

\usepackage{sidecap}

\begin{document}

\mainmatter

\title{Performance analysis of the Kahan-enhanced scalar product on current multicore processors}

\titlerunning{Performance analysis of reduction kernels}

\author{J.~Hofmann\textsuperscript{1} \and D.~Fey\textsuperscript{1} \and J.~Eitzinger\textsuperscript{2} \and G.~Hager\textsuperscript{2} \and G.~Wellein\textsuperscript{2}}
\authorrunning{J. Hofmann et al.}

\institute{
   \textsuperscript{1}Chair for Computer Architecture, University Erlangen-Nuremberg\\
   \textsuperscript{2}Erlangen Regional Computing Center (RRZE), University Erlangen-Nuremberg}

\maketitle
\begin{abstract}
  We investigate the performance characteristics of a numerically
  enhanced scalar product (dot) kernel loop that uses the Kahan
  algorithm to compensate for numerical errors, and describe efficient
  SIMD-vectorized  implementations on recent Intel processors. Using low-level
  instruction analysis and the execution-cache-memory (ECM)
  performance model we pinpoint the relevant performance bottlenecks
  for single-core and thread-parallel execution, and predict
  performance and saturation behavior. We show that the
  Kahan-enhanced scalar product comes at almost no additional cost
  compared to the naive (non-Kahan) scalar product if appropriate
  low-level optimizations, notably SIMD vectorization and unrolling,
  are applied. We also investigate the impact of architectural changes
  across four generations of Intel Xeon processors.
\end{abstract}
%
\section{Introduction and related work}

Accumulating finite-precision floating-point numbers in a scalar variable 
is a common operation in
computational science and engineering. The consequences in terms of accuracy 
are inherent to the number representation and have been well known and studied 
for a long time~\cite{Goldberg:1991}. There is a number of summation 
algorithms that enhance accuracy while
maintaining an acceptable throughput~\cite{Linz:1970,Gregory:1972},
of which  Kahan~\cite{Kahan:1965} 
is probably the
most popular one. However, the topic is still subject to active research
\cite{Rump:2008,Zhu:2010,Demmel:2013,Dalton:2014}.
A
straightforward solution to the inherent accuracy problems is
arbitrary-precision floating point arithmetic, which comes at a
significant performance penalty.
Naive summation and arbitrary precision arithmetic are at opposite
ends of a broad spectrum of options, and
balancing performance vs.\ accuracy is a key concern when selecting
a specific solution. 

Naive summation, which simply adds each successive number in sequence
to an accumulator, requires appropriate unrolling for SIMD
vectorization and pipelining. The necessary code transformations are
performed automatically by modern compilers, which results in optimal
in-core performance.  Such a code quickly saturates the memory
bandwidth of modern multi-core CPUs when the data is in memory.

This paper investigates implementations of the scalar
product, a kernel which is relevant in many numerical algorithms. Starting from
an optimal naive implementation it considers scalar and SIMD-vectorized
versions of the Kahan algorithm using various SIMD
instruction set extensions on a range of current Intel processors.
Using an analytic performance model we point out the conditions under
which Kahan comes for free, and we predict the single core performance
in all memory hierarchy levels as well as the scaling behavior across
the cores of a chip.


\section{Performance modeling on the core and chip level}\label{sec:models}

%


%
The \ecmm~\cite{Treibig:2009,hager:cpe,sthw15} is an extension of
the well-known \Rlm~\cite{roofline:2009}. It estimates the
number of CPU cycles required to execute a number of iterations
$n_\mathrm{it}$ of a loop on a single core of a multicore chip. It
considers the time for executing the iterations with data coming from
the L1 cache as well as the time for moving the required cache lines
(CLs) through the cache hierarchy. In the following we will assume
fully inclusive caches, which is appropriate for current Intel architectures. We give a brief overview of the model
here; details can be found in~\cite{sthw15}.


The \ecmm\ considers the
time to execute the instructions of a loop kernel on the processor core, assuming
that there are no cache misses, and the time to transfer data between
its initial location and the L1 cache. 
The in-core execution time $T_\mathrm{core}$ is determined by the
unit that takes the
most cycles to execute the instructions.
Since data transfers in the memory hierarchy occur in units of cache
lines (CLs), we always consider one cache line's ``worth of work.''
E.g., with a loop
kernel that handles single-precision floating-point arrays with unit
stride, one unit of work is $n_\mathrm{it}=16$~iterations.

The time needed for all data transfers required to execute one work
unit is the ``transfer time.'' We neglect all latency effects, so
the cost for one CL transfer is set by the maximum bandwidth. 
E.g., on the Intel IvyBridge architecture, one CL transfer takes two cycles between
adjacent cache levels. Getting a 64-\byte\ CL from memory to L3 or
back takes $64\,\bytes\cdot f/b_\mathrm S$ cycles, where $f$ is the
CPU clock speed and $b_\mathrm S$ is the memory bandwidth. 
Note that in practice we encounter the problem
that the model is too optimistic for in-memory data sets on some
processors. This can
be corrected by introducing a latency penalty. See Sect.~\ref{sec:impl}
for details.

The in-core execution and transfer times must be put together
to arrive at a prediction of single-thread execution time.
If $T_\mathrm{data}$
is the transfer time,
$T_\mathrm{OL}$ is the part of the core execution that overlaps with
the transfer time, and $T_\mathrm{nOL}$ is the part that does not, then
\bq
T_\mathrm{core} =  \max\left(T_\mathrm{nOL},T_\mathrm{OL}\right)\quad\mbox{and}\quad
T_\mathrm{ECM} =  \max(T_\mathrm{nOL}+T_\mathrm{data},T_\mathrm{OL})\label{eq:T}\eos
\eq
The model assumes that 
(i) core cycles in which loads are retired do not overlap with any
other data transfer in the memory hierarchy, but all other in-core
cycles (including pipeline bubbles) do, and (ii) the transfer times
up to the L1 cache are mutually non-overlapping.
A shorthand notation is used to summarize the relevant information
about the cycle times that comprise the model for a loop: We write
the model as \ecm{T_\mathrm{OL}}{T_\mathrm{nOL}}{T_\mathrm{L1L2}}{T_\mathrm{L2L3}}{T_\mathrm{L3Mem}}{}, where $T_\mathrm{nOL}$ and
$T_\mathrm{OL}$ are as defined above, and the
other quantities are the data transfer times between adjacent memory
hierarchy levels. 
Cycle predictions for data sets fitting into any given memory level can be
calculated from this  by adding up the
appropriate contributions from $T_\mathrm{data}$ and $T_\mathrm{nOL}$
and applying (\ref{eq:T}).
For instance, if the \ecmm\ reads \ecm{2}{4}{4}{4}{9}{\cycles}, the prediction
for L2 cache will be $\max\left(2,4+4\right)\,\cycles=8\,\cycles$.
As a shorthand notation
for predictions we use a similar format but with ``$\epsep$''
as the delimiter. For the above example
this would read as
$T_\mathrm{ECM}=\ecmp{4}{8}{12}{21}{\cycles}$. Converting from
time (cycles) to performance is done by dividing the work $W$
(e.g., \flops) by the runtime: $P=W/T_\mathrm{ECM}$.
If $T_\mathrm{ECM}$ is given in clock cycles but the desired unit of performance is 
\FS, we have to multiply by the clock speed.


We assume that the
single-core performance scales linearly until a bottleneck is
hit. On modern Intel processors the
only bottleneck is the memory bandwidth, which means that an upper
performance limit is given by the Roof\/line prediction for memory-bound
execution: $P_\mathrm{BW} = I\cdot b_\mathrm S$,
where $I$ is  the computational intensity of the loop code.
The performance scaling for $n$ cores is thus described by
$P(n)=\min\left(nP_\mathrm{ECM}^\mathrm{mem},I\cdot b_\mathrm S\right)$
if $P_\mathrm{ECM}^\mathrm{mem}$ is the \ecmm\ prediction for
data in main memory. 
The performance will saturate at $n_\mathrm{S}=\left\lceil T_\mathrm{ECM}^\mathrm{mem}/T_\mathrm{L3Mem}\right\rceil$ cores.
In the following section we will use the \ecmm\ to describe performance properties
of different dot implementations.

\section{Optimal implementations and performance models for dot}\label{sec:impl}
%
Table~\ref{tab:arch} gives an overview of the relevant architectural details of
the four generations of Intel Xeon processors used in this work.
\begin{table}[tb]
\renewcommand{\arraystretch}{1.2}\centering
\begin{tabular}{l@{\hspace{5pt}}c@{\hspace{5pt}}c@{\hspace{5pt}}c@{\hspace{5pt}}c}
        \hline
	Microarchitecture         &SandyBridge-EP& IvyBridge-EP  & Haswell-EP   & Broadwell-D \\
        Shorthand                 & SNB & IVB & HSW & BDW \\
	    Xeon Model            &E5-2680        &E5-2690 v2     &E5-2695 v3    &D-1540     \\
	    Year                      &03/2012             &09/2013             &09/2014            &03/2015         \\
        \hline
	    Clock speed (fixed)       &2.7\,\GHZ            &2.2\,\GHZ            &2.3\,\GHZ            &1.8\,\GHZ   \\
        Cores/Threads             &8/16                &10/20               &14/28               &8/16    \\
        \hline
        Load/Store \makebox[0pt][l]{throughput per cycle} &                            &                 &                 \\
         ~~~AVX(2)                &1\,LD\,\&\,1/2\,ST         &1\,LD\,\&\,1/2\,ST         &2\,LD\,\&\,1\,ST & 2\,LD\,\&\,1\,ST  \\
         ~~~SSE/scalar            &2\,LD\,$\|$\,1\,LD \& 1\,ST         &2\,LD\,$\|$\,1\,LD\,\&\,1\,ST   &2\,LD\,\&\,1\,ST\,&2\,LD\,\&\,1\,ST   \\
        L1 port width             &2$\times$16+1$\times$16\,\BYTE  &2$\times$16+1$\times$16\,\BYTE    &2$\times$32+1$\times$32\,\BYTE &2$\times$32+1$\times$32\,\BYTE  \\
        ADD throughput     &1 / cy              &1 / cy              &1 / cy &1 / cy   \\
        MUL throughput     &1 / cy              &1 / cy              &2 / cy &2 / cy   \\
        FMA throughput     &n/a                 &n/a                 &2 / cy &2 / cy   \\
        \hline
        L2-L1 data bus   & 32\,\BYTE         & 32\,\BYTE         & 64\,\BYTE         & 64\,\BYTE         \\
        L3-L2 data bus   & 32\,\BYTE         & 32\,\BYTE         & 32\,\BYTE         & 32\,\BYTE         \\
	    LLC size                  &20\,\MiB              &25\,\MiB              &35\,\MiB       &12\,\MiB   \\
	    Main memory  &4$\times$DDR3-1600&4$\times$DDR3-1866&4$\times$DDR4-2133 &4$\times$DDR4-2133  \\
        Peak memory BW     &51.2\,\GBS          &51.2\,\GBS          &68.3\,\GBS         &34.1\,\GBS \\
        Load-only BW  &43.6\,\GBS\ (85\%) &46.1\,\GBS\ (90\%) &60.6\,\GBS\ (89\%) & 33\,\GBS\ (95\%) \\
        $T_\mathrm{L3Mem}$ per CL  & 3.96\,\cycles   & 3.05\,\cycles   & 2.43\,\cycles  & 3.49\,\cycles \\
        \hline
	\end{tabular}\\[3mm]
    \caption{Test machine specifications and micro-architectural features (one socket). The cache line length is 64\,\bytes\ in all cases.
    The SIMD register width is 16\,\bytes\ for SSE and 32\,\bytes for AVX.}
    \label{tab:arch}
\end{table}
The CPUs were released in successive years between 2012 and 2015.
Intel Haswell-EP marks the big micro-architectural change, with a new SIMD
instruction set extension (AVX2) and several fused multiply-add instructions (FMA3).
There are also notable improvements in the memory hierarchy: The
access path width of load/store units was widened from 16\,\bytes\ to 32\,\bytes,
and the bus width
between the L2 and the L1 cache was enlarged from 32\,\bytes\ to 64\,\bytes.
The Broadwell chip is a very recent power-efficient ``Xeon D'' variant. All results
for Broadwell are preliminary since we only had access to a pre-release
version of the chip.

%
%

We first discuss variants for dot in single precision (SP) for the
Intel IvyBridge microarchitecture. The differences to double precision (DP)
and the impact of architectural changes are
covered in Sect.~\ref{sec:double}.
To eliminate variations introduced by compilers we implemented all kernels directly 
in assembly language using the \verb.likwid-bench. microbenchmarking 
framework~\cite{Treibig:2011:3}.

\subsubsection{Naive scalar product}
The naive scalar product in single precision
serves as the baseline (see Fig.~\ref{fig:listings}a).
\begin{figure}[tb]
(a)\begin{minipage}[t]{0.43\linewidth}
\begin{lstlisting}
  float sum = 0.0;
    
  for (int i=0; i<n; i++) {
     sum = sum + a[i] * b[i]
  }
  \end{lstlisting}
\end{minipage}\hfill
(b)\begin{minipage}[t]{0.46\linewidth}    
\begin{lstlisting}
    float sum = 0.0;
    float c = 0.0;
    for (int i=0; i<N; ++i) {
        float prod = a[i]*b[i];
        float y = prod-c;
        float t = sum+y;
        c = (t-sum)-y;
        sum = t;
    }
\end{lstlisting}
\end{minipage}
  \caption{\label{fig:listings}(a) Naive scalar product code in single precision.
    (b) Kahan-compensated scalar product code.}  
\end{figure}
Sufficient unrolling must be applied to hide the ADD pipeline latency
for the recursive update on the accumulation register and to apply
SIMD vectorization. Both optimizations introduce partial sums and are
therefore not compatible with the C standard as the order of
non-associative operations is changed.  With higher optimization levels the
current Intel compiler (version 15.0.2) is able to generate 
optimal code.  Note that partial sums usually improve the accuracy of
the result \cite{Dalton:2014}.

This kernel is limited by the throughput of the LOAD unit on
the IVB architecture (see Table~\ref{tab:arch}). 
Two AVX loads per vector (\verb.a. and \verb.b.) are required to cover
one unit of work (16 scalar loop iterations), resulting
in $T_\mathrm{nOL}=4\,\cycles$. The overlapping 
part is $T_\mathrm{nOL}=2\,\cycles$ since two MULT and two ADD instructions
must be executed. Data transfers between cache levels require two cycles
per CL, so that $T_\mathrm{L1L2}=T_\mathrm{L2L3}=4\,\cycles$.

For $T_\mathrm{L3Mem}$ we calculate the number of cycles per CL transfer
from the maximum memory bandwidth and the clock speed (last row in Table~\ref{tab:arch})
and arrive at $T_\mathrm{L3Mem}=6.1\,\cycles$. The full 
\ecmm\ thus reads \ecm{2}{4}{4}{4}{6.1}{\cycles}. On newer Intel
chips (notably IVB and HSW) unknown peculiarities in the 
design of the Uncore lead to extra latency penalties per cache line
from memory. We take these deviations into account by introducing 
a penalty parameter that is fixed empirically. This parameter
is an additive contribution to $T_\mathrm{L3Mem}$,
so that the final model is \ecm{2}{4}{4}{4}{6.1+2.9}{\cycles},
leading to a runtime prediction of \ecmp{4}{8}{12}{18.1+2.9}{\cycles}. 
At a clock speed of 2.2\,\GHZ\ the expected serial performance is
thus
\bq
P=\frac{16\,\updates\cdot 2.2\,\GCS}{\ecmp{4}{8}{12}{18.1+2.9}{\cycles}} =
\ecmp{8.80}{4.40}{2.93}{1.68}{\GUS}\eos
\eq
We choose an ``\update'' (two\,\flops) as the basic unit of work to make performance
results for different implementations comparable.
The predicted saturation point is at
$n_\mathrm S=\left\lceil (18.1+2.9)/6.1\right\rceil=4$ cores.
Note that the maximum memory bandwidth has to be taken into account
for the saturation point, so we divide by 6.1\,\cycles.
The Roof\/line ``light speed,'' i.e., the memory bandwidth-limited saturated
performance, can be calculated from the computational intensity of one \update\
per eight\,\bytes:
$P_\mathrm{BW}=\left(1\,\update/8\,\BYTE\right)\cdot b_\mathrm S=5.76\,\GUS$. 

 All versions of the
enhanced scalar product described in the next section will be compared
to the optimal naive implementation.
\subsubsection{Kahan-enhanced scalar product on IvyBridge}
Figure~\ref{fig:listings}b shows the implementation of the Kahan algorithm for dot.
Compilers have problems with this loop code for two reasons: First, the compiler
detects (correctly) a loop-carried dependency on $c$, which prohibits SIMD
vectorization and modulo unrolling. Second, the compiler may recognize that,
arithmetically, $c$ is always equal to zero. With high optimization levels
it may thus reduce the code to the naive scalar product, defeating the purpose
of the Kahan algorithm. This is the reason why we use hand-coded assembly
throughout this work.

One iteration comprises one multiplication, four additions or
subtractions, and two loads. The
bottleneck on the IVB core level is thus the ADD unit
(ADD and SUB are handled by the same pipeline). In the
following we construct the \ecmm\ for scalar, SSE, and AVX versions of
the Kahan loop.  Independent of vectorization we always establish
proper modulo unrolling for best pipeline utilization.
%
%

\emph{Scalar implementation.} In scalar mode, one unit of work amounts to $16\times 4=64$
instructions in the ADD unit, resulting in
$T_\mathrm{OL}=64\,\cycles$. Since two scalar loads can be executed
per cycle on the IVB core, the 32 loads lead to
$T_\mathrm{nOL}=16\,\cycles$.  The contributions from in-cache and
memory transfers are the same as for the naive variant above, so the
complete \ecmm\ is \ecm{64}{16}{4}{4}{6.1+2.9}{\cycles}, and the
runtime prediction is \ecmp{64}{64}{64}{64}{\cycles}. According to the
model the scalar variant should not be able to saturate the memory
bandwidth using all cores on the ten-core chip, since $n_\mathrm{S}=\left\lceil
64/6.1\right\rceil=11$\,cores.
The analysis shows that the scalar variant of Kahan is limited by the instruction
throughput, specifically on the ADD pipeline, regardless of where the data
resides. We thus expect the same performance $P=16\cdot 2.2/64\,\GUS=0.55\,\GUS$
in all memory hierarchy levels for single-threaded execution, and close to perfect
scalability across the cores of the chip.

%
\emph{SSE implementation.} SSE uses 16-\byte\ wide registers, and all instructions required for
the Kahan algorithm exist in SSE variants, so the overall number of instructions
is reduced by a factor of four compared to the scalar version, but the same
throughput limits apply for the ADD and the LOAD unit. This leads to
an \ecmm\ of \ecm{16}{4}{4}{4}{6.1+2.9}{\cycles} and a prediction
of \ecmp{16}{16}{16}{18.1+2.9}{\cycles}, which yields
$P=\ecmp{2.20}{2.20}{2.20}{1.68}{\GUS}$. 
The SSE code is limited by the instruction throughput up to the L3
cache since all data transfer contributions can be overlapped
with the ADD instructions. The optimal 4$\times$
speed-up of SSE is thus observed in this case. For data in main memory
the speed-up is just about $64/21\approx 3${}$\times$, and
the single-core performance and saturation behavior are identical
to the naive scalar product.

%
\emph{AVX implementation.} AVX further reduces the runtime for the ADD operations by a factor of
two, so $T_\mathrm{OL}=8\,\cycles$. Although the number of LOAD
instructions is also cut in half, the non-overlapping time
$T_\mathrm{nOL}$ does not change, because the two LOAD ports of
the L1 cache are only 16\,\bytes\ wide. Therefore only one
LOAD instruction can be retired per cycle. 
The complete ECM model is \ecm{8}{4}{4}{4}{6.1+2.9}{\cycles}, the
runtime prediction is \ecmp{8}{8}{12}{18.1+2.9}{\cycles} (leading to
$P=\ecmp{4.40}{4.40}{2.93}{1.68}{\GUS}$), and the saturation behavior
is the same as for the SSE variant of Kahan and the naive scalar
product.
The AVX code is limited by the instruction throughput up to the L2
cache, and the full 2$\times$ advantage versus SSE can be observed in this
case. Starting from L3 there is a slight impact on runtime by data transfers,
leading to a reduced speed-up of 1.3$\times$ in L3 and none at all
in main memory. Again the saturation behavior is expected
between three and four cores.

The conclusion from this analysis is that there is no expected
performance difference for in-memory working sets between the naive
scalar product and the Kahan version if any kind of vectorization is
applied to Kahan. With AVX, Kahan comes for free even in the L3 or the
L2 cache. Only for in-L1 data we expect a 2$\times$ slowdown for
Kahan versus the naive version even with the best possible code.
%
\subsubsection{Influence of processor architecture}\label{sec:arch}
\begin{table}[tb]
  \renewcommand{\arraystretch}{1.2}
  \centering
  \begin{tabular}{lccc}
            & \ecmm\ [\cycles]                     & Prediction [\cycles/CL]            & Pred. performance [\GUS] \\\hline      
    SNB     & \ecm{8}{4}{4}{4}{7.9+5.1}{}          & \ecmp{8}{8}{12}{19.9+5.1}{}        & \ecmp{5.40}{5.40}{3.60}{1.73}{}\\
    IVB     & \ecm{8}{4}{4}{4}{6.1+2.9}{}          & \ecmp{8}{8}{12}{18.1+2.9}{}        & \ecmp{4.40}{4.40}{2.93}{1.68}{}\\
    HSW~~~  & \ecm{8}{\textbf 2}{\textbf 2}{5.54}{4.9+11.1}{}      & \ecmp{8}{8}{9.54}{14.44+11.1}{}    & \ecmp{4.60}{4.60}{3.86}{1.44}{}\\
    BDW     & \ecm{8}{2}{2}{\textbf 4}{7+\textbf 1}{}              & \ecmp{8}{8}{8}{15+1}{}             & \ecmp{3.60}{3.60}{3.60}{1.8}{}
  \end{tabular}
  \caption{\label{tab:ecm_arch}Comparison of the \ecmm\ for optimal AVX implementations across the
    multicore Xeon CPUs in the testbed (see Table~\ref{tab:arch}). The consequences of relevant architectural changes
    to the preceding generation are highlighted.}
\end{table}
In this section we compare the model-based analysis across four
generations of Intel CPUs: SandyBridge-EP (SNB), IvyBridge-EP (IVB),
Haswell-EP (HSW), and Broadwell (BDW, in a power-efficient ``Xeon D''
variant).  This covers four Intel Xeon microarchitectures over a time
of three years and involves one major architectural step (from IVB
to HSW). We always consider the optimal AVX code for the comparisons. 
There is no major change expected between SNB and IVB, since no
dot-relevant hardware features were added.  All observed performance
differences are thus rooted in the clock speed and memory
bandwidth (first row in Table~\ref{tab:ecm_arch}).
Note that despite the lower
memory bandwidth of the SNB test system compared to IVB, the in-memory
performance is higher due to the faster clock speed of SNB.
The HSW microarchitecture has new features which influence dot
performance: It can sustain two AVX loads and one AVX store per cycle,
effectively doubling LOAD/STORE throughput. In addition, the L1-L2 bus
width was doubled, allowing for a full CL transfer per cycle. These
changes result in $T_\mathrm{nOL}=2\,\cycles$ and
$T_\mathrm{L1L2}=2\,\cycles$ (third row in Table~\ref{tab:ecm_arch}).
Here we encounter the
peculiarity that HSW lowers the Uncore clock speed if
only a single core is used. This is the reason why the
$T_\mathrm{L2L3}$ contribution is $5.54\,\cycles$ instead of
$4\,\cycles$. 
The BDW architecture has introduced no relevant changes, but it does not
show the Uncore slowdown like HSW (fourth row in Table~\ref{tab:ecm_arch}).\footnote{Note that our test system was
  a pre-release Xeon D; production systems and mainstream Xeon Broadwell
  chips may show a different behavior.}
It is interesting that BDW only requires half a cycle of latency penalty
per CL in memory, so the uncorrected \ecmm\ works very well already. BDW performance
is insensitive to data transfers up to the L3 cache.
\begin{SCfigure}[0.9][tb]
\includegraphics*[width=0.5\linewidth]{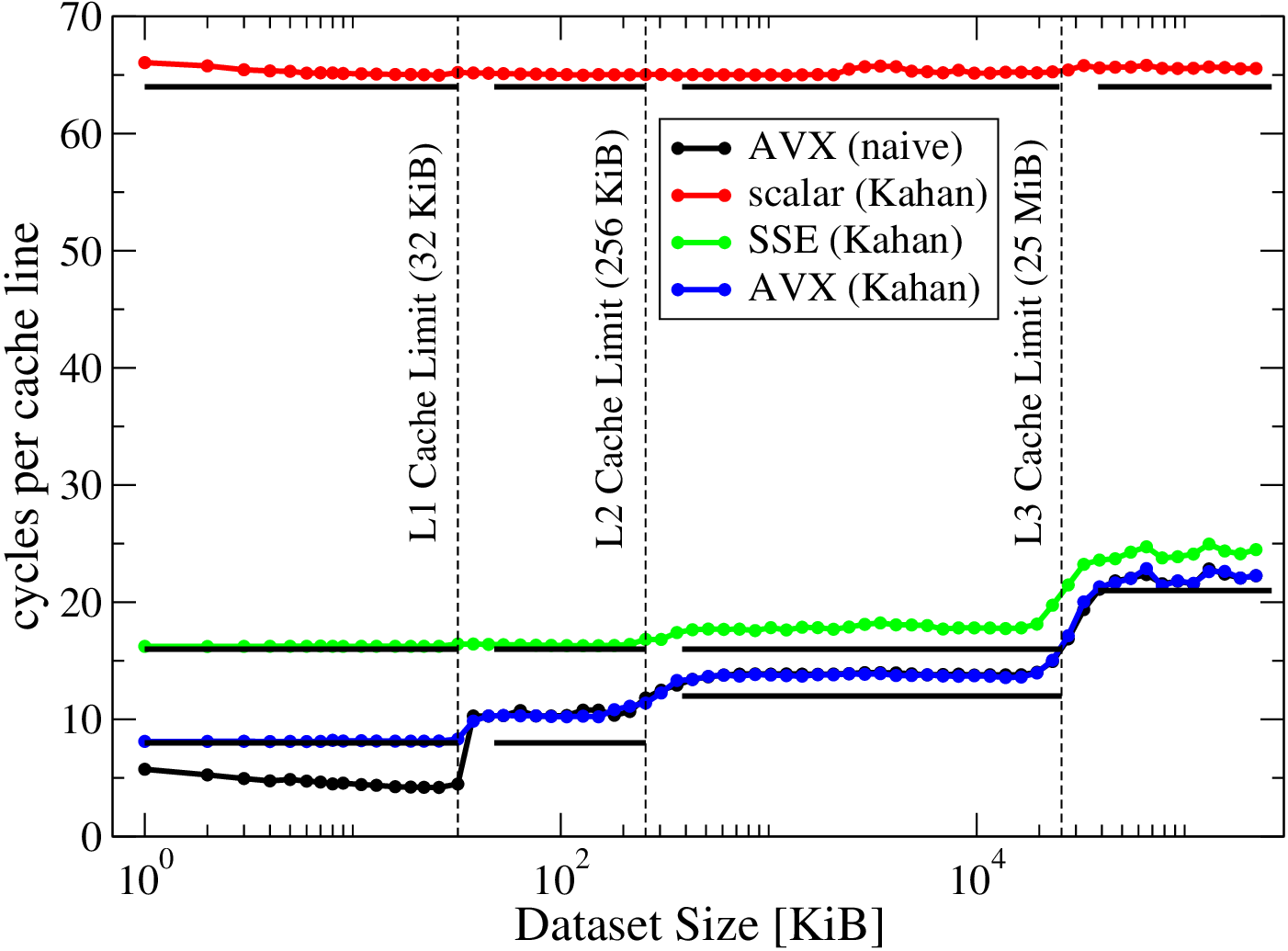}
\caption{\label{fig:kahan_ivy_core}Single-core cycles per CL vs.\ 
  data set size for various implementations of the Kahan scalar
  product and the AVX version of the naive scalar product in SP on
  IVB. The horizontal lines represent the \ecmm\
  predictions for scalar (top), SSE (middle), and AVX (bottom)
  Kahan variants.\bigskip}
\end{SCfigure}

\subsubsection{Double vs.\ single precision}\label{sec:double}
The model prediction in terms of cycles per CL does not change for the
SIMD variants of Kahan when going from SP to DP, but one CL update
represents twice as much useful work (scalar iterations) in the SP
case.  However, the penalty for going from SIMD to scalar is only half
as big as for SP, since the scalar register width is eight bytes
instead of four. The ECM model for the DP scalar version of the Kahan
dot on IVB is \ecm{32}{8}{4}{4}{6.1+2.9}{\cycles} and the according
runtime prediction is \ecmp{32}{32}{32}{32}{\cycles}, with
$P=0.55\,\GUS$. The reduced
cycle count (32 instead of 64) for DP leads to saturation at a smaller
number of cores for in-memory working sets: $n_\mathrm
S=\left\lceil32/6.1\right\rceil=6$.  Hence, even the scalar DP variant
of Kahan exerts sufficient pressure on the memory interface to reach
saturation. The saturated DP performance according to the \Rlm\ is
$P_\mathrm{BW}=(1\,\update/16\,\BYTE)\cdot b_\mathrm S=2.88\,\GUS$.

\section{Performance results and model validation}\label{sec:results}
%
%
%
Single-core benchmarking results for single precision
on IVB are shown in Fig.~\ref{fig:kahan_ivy_core}. The
model predicts the overall behavior very well. The naive and the AVX Kahan version
show identical performance in L2 cache and beyond. As predicted there is no
performance drop for the SSE Kahan version from L1 to L2.
Both AVX Kahan and the compiler-generated naive version fall slightly
short of the prediction in L2. This is a general observation with many
loop kernels, and we interpret it as a consequence of the
L2-L1 hardware prefetcher doing a better job in latency hiding for SSE than for
AVX due to the more relaxed timings in the SSE case. Since the details
of prefetching are undisclosed, we have no way to prove or refute
this hypothesis.
Finally, the constant performance of the scalar Kahan variant across all memory
levels is perfectly predicted by the model.

\begin{figure}[tb]
\includegraphics*[width=0.47\linewidth]{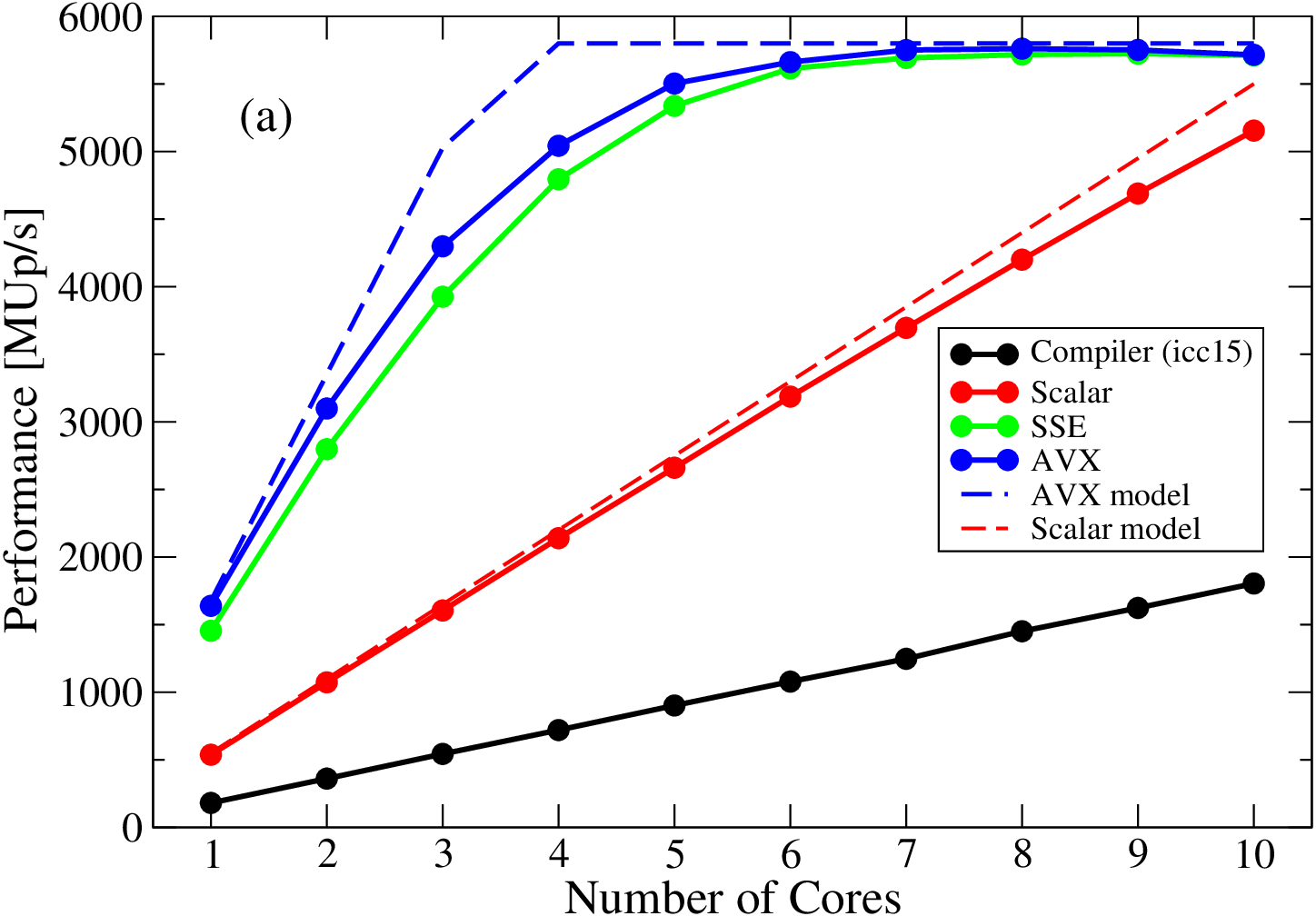}\hfill
\includegraphics*[width=0.47\linewidth]{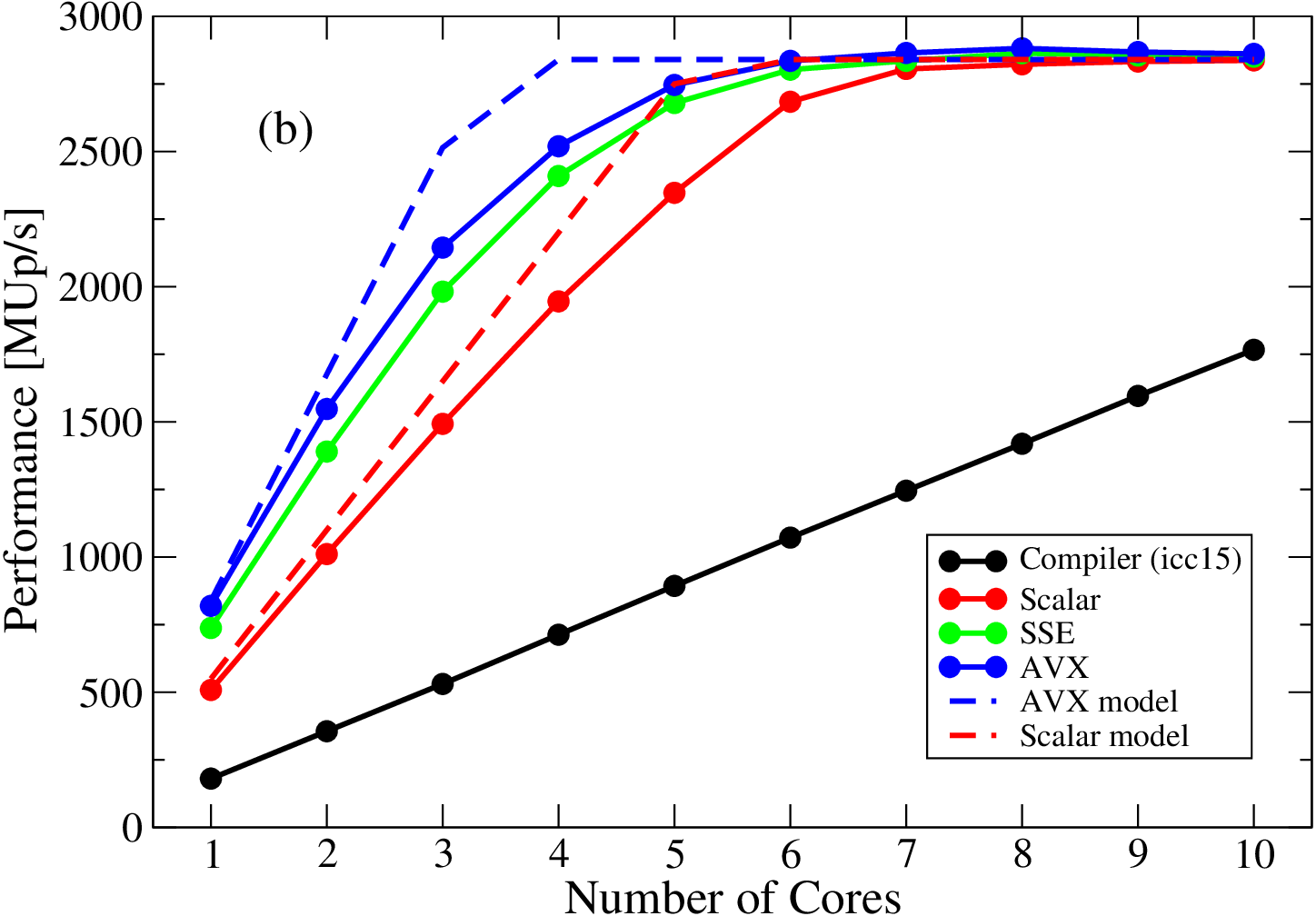}
\caption{\label{fig:kahan_ivy_scaling}In-memory scaling for different
  implementation of the Kahan scalar product on IVB for (a) single
  precision and (b) double precision. Model predictions are shown with
  dashed lines for the scalar and AVX versions.}
\end{figure}

In-memory scaling
results on the chip level are shown in Fig.~\ref{fig:kahan_ivy_scaling}a. The
dashed lines are the model predictions (for clarity we only show
models for scalar and AVX). As anticipated via the \ecmm, 
the scalar version cannot saturate the memory bandwidth even if all cores
are used. Since any code that is 
able to saturate the bandwidth is ``perfect,'' any kind of vectorization
will make the Kahan algorithm as fast as the naive scalar product. Note, however,
that on a CPU with a faster clock speed or more cores saturation will
be easily achieved even with scalar code. This effect illustrates the
general observation that more
parallelism can ``heal'' low single-core performance.
For comparison we also show the compiler-generated variant of Kahan.
As described earlier, the code is devastatingly slow since the compiler
cannot resolve the loop-carried dependency.

%
%
The only relevant difference between DP and SP is the smaller
performance penalty of the scalar variant for DP (see
Fig.~\ref{fig:kahan_ivy_scaling}b), which leads to strong
saturation at about six cores as predicted by the model.

%
\begin{figure}[tb]
\hspace*{\fill}\includegraphics*[width=0.48\linewidth]{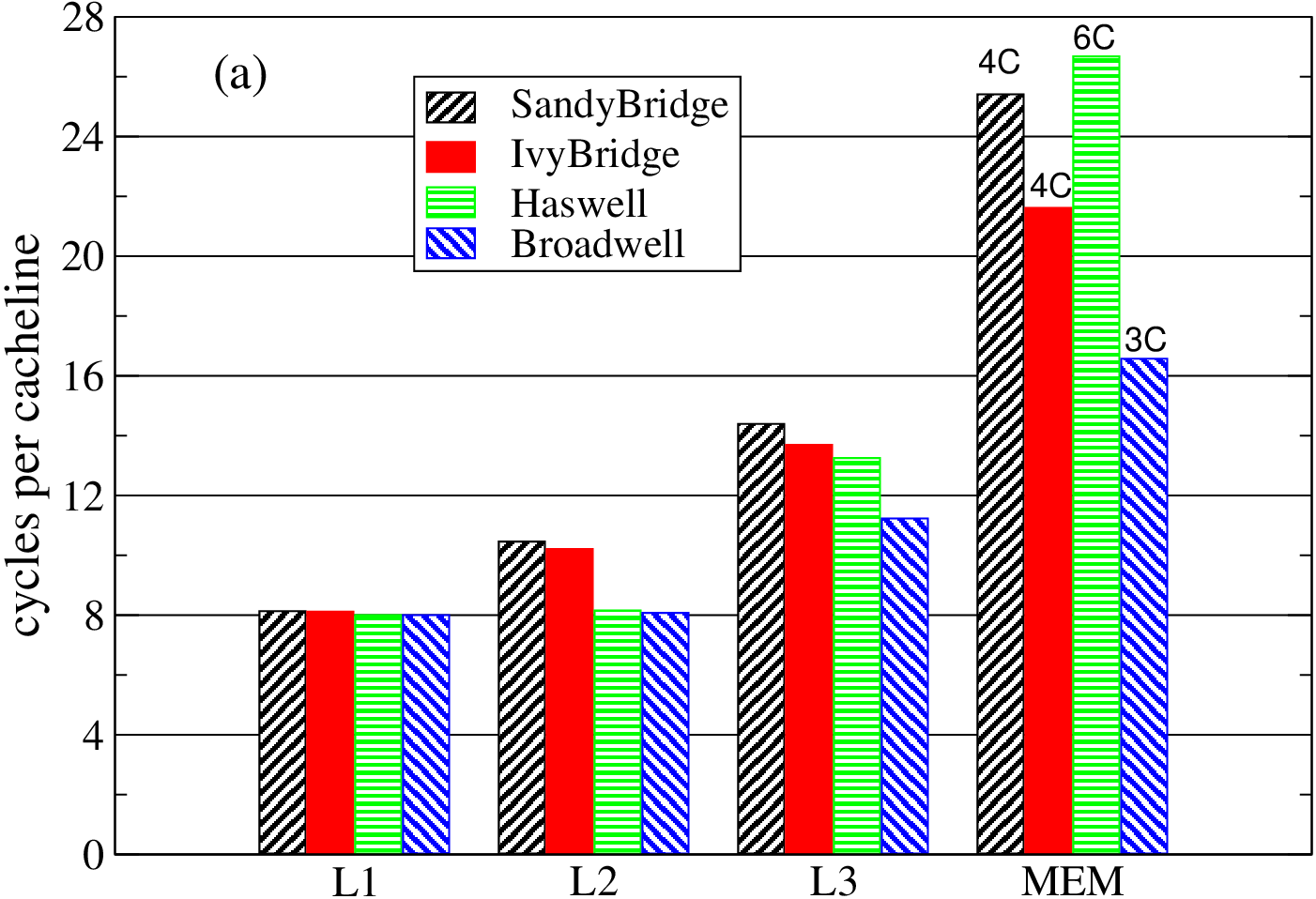}\hfill
\includegraphics*[width=0.3\linewidth]{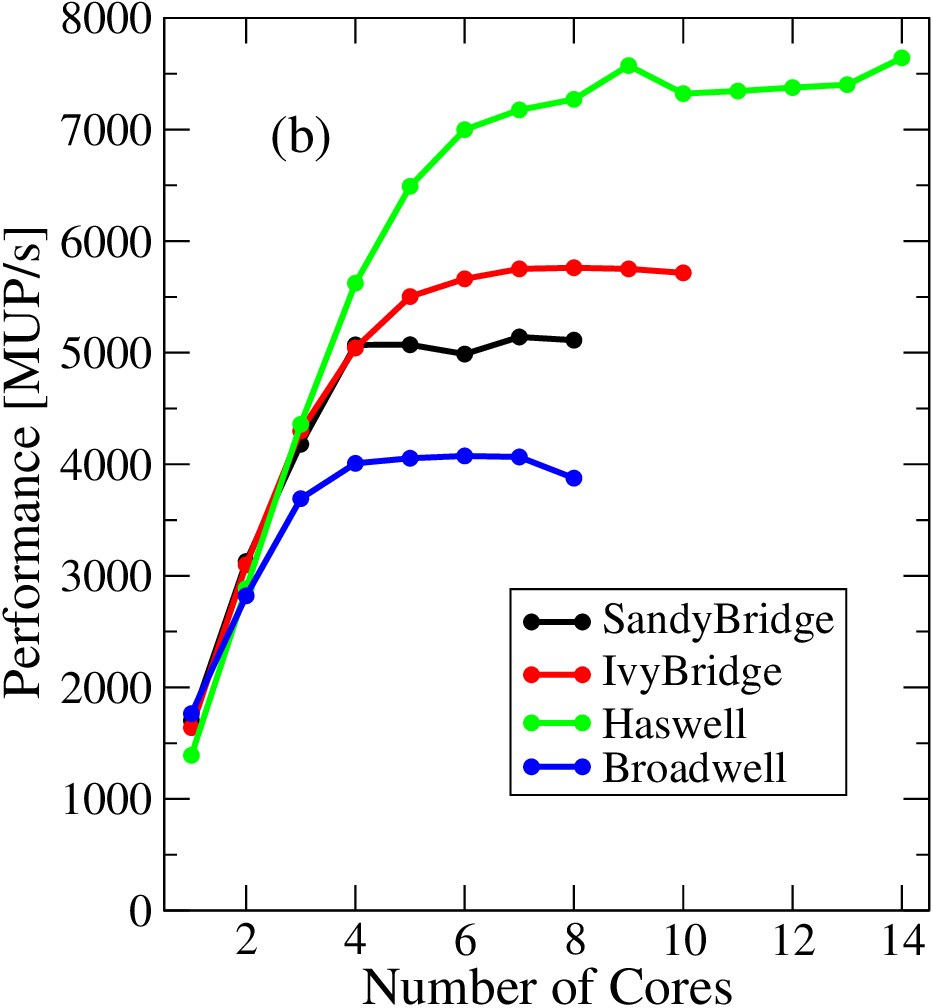}\hspace*{\fill}
\caption{\label{fig:kahan_single_core_archs}Comparison between four
  Intel Xeon multi-core architectures using the single-precision
  AVX Kahan scalar product:
  (a) Measured single-core runtime in cycles per CL in
  different memory hierarchy levels. The saturation point $n_\mathrm S$ is
  indicated above the bars for the memory-bound case. (b) 
  Measured performance scaling with in-memory working set.}
\end{figure}
In order to compare different
architectures, we show single-core data for the AVX-vectorized Kahan scalar
product in Fig.~\ref{fig:kahan_single_core_archs}a.
Since we report runtime in cycles per CL, the influence of
different clock speeds and memory bandwidths is only visible
for the in-memory case.
In the L1 cache all processors show the same runtime, because the
architectural improvements with HSW and BDW do not address the
bottleneck of the algorithm at hand (the ADD throughput). In L2 and
L3, HSW and BDW show higher performance than the previous generations
due to their doubled L2-L1 bandwidth and LOAD throughput.  The step
from L2 to L3 is important, since it marks the transition to the
Uncore, which is a shared resource across the cores. Although a
notable improvement is seen in L3 with each new architecture, we
observe efficiency issues in the Uncore on IVB and HSW that prevent
those CPUs from attaining the expected performance. In memory, HSW is
a significant step back in terms of single-core performance due to a
large latency penalty. BDW seems to have corrected those issues, but
we must stress again that these observations were made on an
eight-core single-socket ``Xeon D'' chip, and it is unclear if the to
be released multi-socket variants with larger core counts can live up
to the expectations raised here. It is also worth emphasizing that, as
already mentioned above, in practice any code that can saturate the
memory bandwidth is ``good enough.''
Figure~\ref{fig:kahan_single_core_archs}b shows the in-memory
performance scaling for all four architectures with the AVX
Kahan scalar product: The differences in saturated performance indeed
reflect the differences in saturated memory bandwidth. Again, and
certainly as expected from the model, vectorization makes the Kahan
algorithm come for free.

There is one additional optimization on HSW and BDW that we have not
mentioned yet. The two FMA units can theoretically increase the ADD
throughput by a factor of two. Both units can execute FMA and MULT
instructions, but only one of them can handle stand-alone ADDs. This
is not a problem with hand-crafted assembly since one can endow an FMA
instruction with a unit multiplicand to act like an ADD.  The downside
is that the FMA instruction has a higher
latency of five cycles (ADD only has three) and therefore requires
deeper unrolling to hide the pipeline latency. Both architectures hence run
out of registers and only achieve a 20\% speed-up from FMA with
data in L1, and no noticeable improvement
beyond L1. 
\section{Conclusion}\label{sec:conc}
We have investigated the performance of naive and Kahan-enhanced
variants of the scalar product on a range of recent Intel multicore chips. 
Using the \ecmm\ the single-core performance in all
memory hierarchy levels and the multi-core scaling for in-memory data
were accurately described.  The most important result
is that even the single-threaded Kahan algorithm comes with no performance penalties on all
standard multicore architectures under investigation in the L2 cache, the L3 cache, and
in memory if implemented optimally. Depending on the particular
architecture and whether single or double precision is used, even
scalar code may achieve bandwidth saturation in memory when using multiple threads.
Performance improvements between successive generations
of Intel CPUs could be attributed to specific architectural
advancements, such as increased LOAD throughput on Haswell
or a more efficient Uncore on Broadwell.

We emphasize that the approach and insights described here for the special
case of the Kahan scalar product can serve as a blueprint for other
load-dominated streaming kernels.


\subsubsection{Acknowledgement}

We thank Intel Germany for providing an early access Broadwell test
system.  This work was partially funded by BMBF under grant 01IH13009A
(project FEPA), and by the Competence Network for Scientific High
Performance Computing in Bavaria (KONWIHR).

\bibliographystyle{splncs}
\bibliography{publications}
\end{document}